\documentclass[preprint2]{aastex62}

\usepackage{natbib}

\usepackage{amsmath}
\usepackage{hyperref}
\usepackage{listings}
\usepackage{color}
\usepackage{float}
\usepackage{xspace} 

\newcommand{\kepler}{\textsl{Kepler}\xspace}
\newcommand{\spitzer}{\textsl{Spitzer}\xspace}

\begin{document}

\title{Possible Bright Starspots on TRAPPIST-1}

\author[0000-0003-2528-3409]{Brett M. Morris}
\affiliation{Astronomy Department, University of Washington, Seattle, WA 98195, USA}

\author{Eric Agol}
\altaffiliation{Guggenheim Fellow}
\affiliation{Astronomy Department and Virtual Planetary Laboratory, University of Washington, Seattle, WA 98195, USA}
\author{James R. A. Davenport}
\altaffiliation{DIRAC Fellow}
\altaffiliation{NSF Astronomy and Astrophysics Postdoctoral Fellow}
\affiliation{Astronomy Department, University of Washington, Seattle, WA 98195, USA}
\affiliation{Department of Physics \& Astronomy, Western Washington University, 516 High St., Bellingham, WA 98225, USA}

\author{Suzanne L. Hawley}
\affiliation{Astronomy Department, University of Washington, Seattle, WA 98195, USA}

\email{bmmorris@uw.edu}

\begin{abstract}
The M8V star TRAPPIST-1 hosts seven roughly Earth-sized planets and is a promising target for exoplanet characterization.  \kepler/K2 Campaign 12 observations of TRAPPIST-1 in the optical show an apparent rotational modulation with a 3.3 day period, though that rotational signal is not readily detected in the \spitzer light curve at 4.5 $\mu$m. If the rotational modulation is due to starspots, persistent dark spots can be excluded from the lack of photometric variability in the \spitzer light curve. We construct a photometric model for rotational modulation due to photospheric bright spots on TRAPPIST-1 which is consistent with both the \kepler and \spitzer light curves. The maximum-likelihood model with three spots has typical spot sizes of $R_\mathrm{spot}/R_\star \approx 0.004$ at temperature $T_\mathrm{spot} \gtrsim 5300 \pm 200$ K. We also find that large flares are observed more often when the brightest spot is facing the observer, suggesting a correlation between the position of the bright spots and flare events. In addition, these flares may occur preferentially when the spots are increasing in brightness, which suggests that the 3.3 d periodicity may not be a rotational signal, but rather a characteristic timescale of active regions.
\end{abstract}


\object{TRAPPIST-1}

\section{Introduction}

TRAPPIST-1 is an M8V star and host to at least seven Earth-sized planets \citep{Gillon2016,Gillon2017}. It was observed with \spitzer at the IRAC-2 4.5 $\mu$m band for 20 days \citep{Gillon2017,Delrez2018}, and 76 days later it was observed by NASA's \kepler spacecraft during K2 Campaign 12 for another 79 days \citet{luger2017everest}.  The K2 observations of TRAPPIST-1 show quasi-periodic flux modulations with a period of 3.3 days \citep{Luger2017, Delrez2018}, which has been interpreted as a signal of stellar rotation as starspots rotate into and out of view \citep{Vida2017}.  Modeling these variations faces the vexing computational problem of fitting stellar rotational flux modulation with a forward model \citep{Aigrain2015, Davenport2015}. One challenge in modeling rotational modulation is that often a light curve can be fit equally well with a few bright spots \textit{or} with a few dark spots. This degeneracy has been referred to as the ``zebra effect''  \citep{Pettersen1992}: is the star bright with a few dark spots, or dark with a few bright spots?

Although star spots are usually dark due to magnetic field pressure balancing thermal pressure, one star with a mostly dark surface and a few bright regions is \textit{T Tauri} star LkCa 4 \citep{Gully2017}. Near-infrared spectroscopy of LkCa 4 shows a heterogeneous atmosphere --- 80\%  of the star has $T_{\mathrm{eff}} \sim 2800$ K, and the other 20\% has $T_{\mathrm{eff}} \sim 4000$ K. Modern Zeeman Doppler Imaging is most sensitive to large scale features in stellar magnetic fields, so it is generally unclear what small-scale magnetic structures may exist on fully-convective stars \citep[see e.g.~][]{Morin2013}.

Another difficulty in interpreting the TRAPPIST-1 brightness variations is whether the periodicity is due to rotation at all.  \cite{Reiners2010} measure a rotational velocity of 6$\pm$2 km sec$^{-1}$, while the 3.3-day period would predict a maximum rotational velocity of 1.8 km sec$^{-1}$.  The spots on TRAPPIST-1 evolve with time, and so it will require long-term monitoring of the 3.3-day periodicity to determine whether it is time-steady, which would argue for a rotational origin, and would call into question the measurement of the rotational velocity.

Another means of studying spots is via occultation during planetary transits, which has been studied in detail on HAT-P-11 \citep{Morris2017a}. To date, no significant spot occultations have been observed in either the \kepler or \spitzer transit light curves of TRAPPIST-1 \citep{Delrez2018} -- this implies that if spots are responsible for the rotational modulation, they must either reside outside of the transit chords of the planets, or be smaller than the size of the planets, which would reduce the probability and amplitude of a spot occultation.

In this paper we present a model for the K2 and Spitzer variations based upon a spot variability
model, and examine its relation to the stellar flares that are frequently seen in both datasets.
In section \ref{rotation_model} we discuss the spot model for stellar variability.  In section
\ref{lightcurve_fits} we compare this model with extant K2 and \spitzer data.  In section
\ref{correlation} we discuss a possible relation between bright flares and the bright spots.
We end with discussion and conclusions.

\section{Rotational variability model} \label{rotation_model}

\subsection{Monochromatic variability}

We compute spot variability at a given wavelength with
a simplified model assuming small spots with radii smaller than
10\% of the stellar radius.
We first integrate the total stellar flux of the unspotted, limb-darkened star,
\begin{equation}
F_{\star, \mathrm{unspotted}} = D^{-2} \int_{0}^{R_\star} 2 \pi r \, I(r) dr,
\end{equation}
where $r$ is the radial coordinate from the center of the star, $D$ is
the distance to the star, $I(r)$ is the specific intensity at radius $r$.  
We model the specific intensity with a quadratic limb-darkening law, using 
the limb-darkening parameters of \citet{Luger2017}: $(u1, u2) = (1.0, -0.04)$. 

We describe  each spot with an ellipse with centroid $\mathbf{r}_i = (x_i, y_i)$
at radial coordinate $r_i = \vert \mathbf{r}_i\vert$.  We
then compute the flux contribution from each spot by computing 
the approximate spot area and spot contrast based on the ratio of Phoenix 
model atmospheres integrated over a given bandpass, which we assume is 
independent of inclination angle of observation.  From the observer's 
perspective, each spot has semimajor axis $R_{spot}$ along the azimuthal
direction,  and semiminor axis $R_{spot} \sqrt{1 - (r_i /R_\star)^2}$ in 
the radial direction, due to foreshortening. Since these spots are
small compared to the stellar radius, $R_{\mathrm{spot}}/R_\star <<1$,  
we adopt a constant limb-darkened contrast for the entire spot,
$c_{ld} = (c-1) I(r_i)/I(0)$, where $c$ is the monochromatic contrast 
in the spot relative to the adjacent photosphere flux;  $c=1$ for an
unspotted region. The integrated  spot flux of each spot is then given by
\begin{equation}
F_{\mathrm{spot}, i} = D^{-2} \pi R_{\mathrm{spot}}^2 c_{ld} I(0) \sqrt{1 - (r_i/R_\star)^2}, 
\end{equation}
where the spotted flux of the star is
\begin{equation}
F_{\star, \mathrm{spotted}} = F_{\star, \mathrm{unspotted}}  + \sum_{i=1}^{N} F_{\mathrm{spot}, i}.
\end{equation}

As the star rotates, the flux of the spots varies due to foreshortening,
limb-darkening, and disappearance behind the edge of the limb. Spots
straddling the limb are ignored until they rotate onto the observer-facing 
hemisphere. This approximation is valid 
for spots that are small compared to the stellar radius, and small compared 
to the scale of limb-darkening variation across the stellar disk. We then 
compute the flux observed as the star rotates, by rotating the positions 
of the spots and recalculating the fluxes.

We find the posterior uncertainties on spot positions (latitude and longitude), radii and contrast in the \kepler bandpass with Markov Chain Monte Carlo \citep{Foreman-Mackey2013}. 

\subsection{Spot contrasts in \kepler and \spitzer}

\begin{figure}
\begin{center}
\includegraphics[scale=0.6]{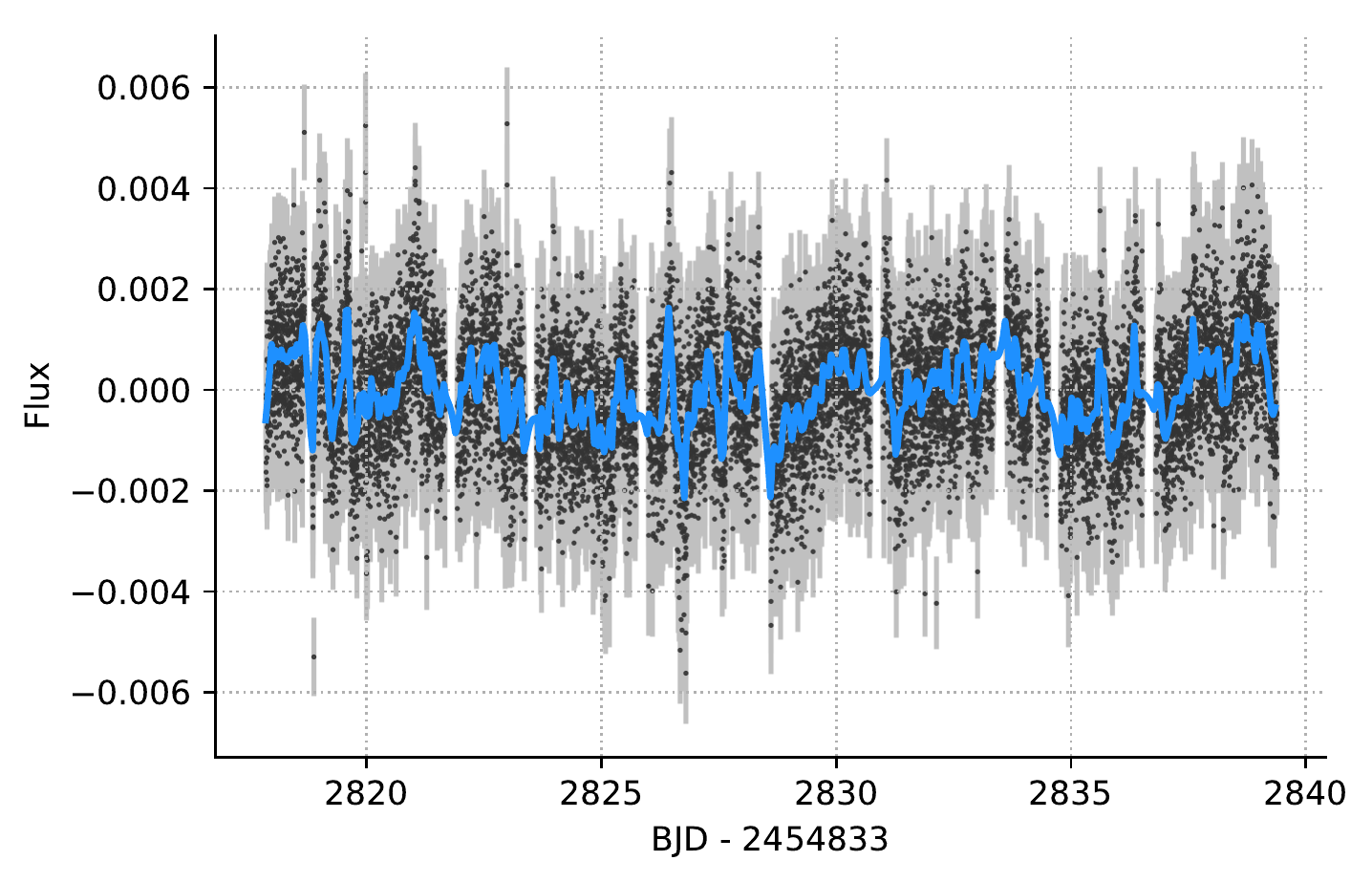}
\includegraphics[scale=0.5]{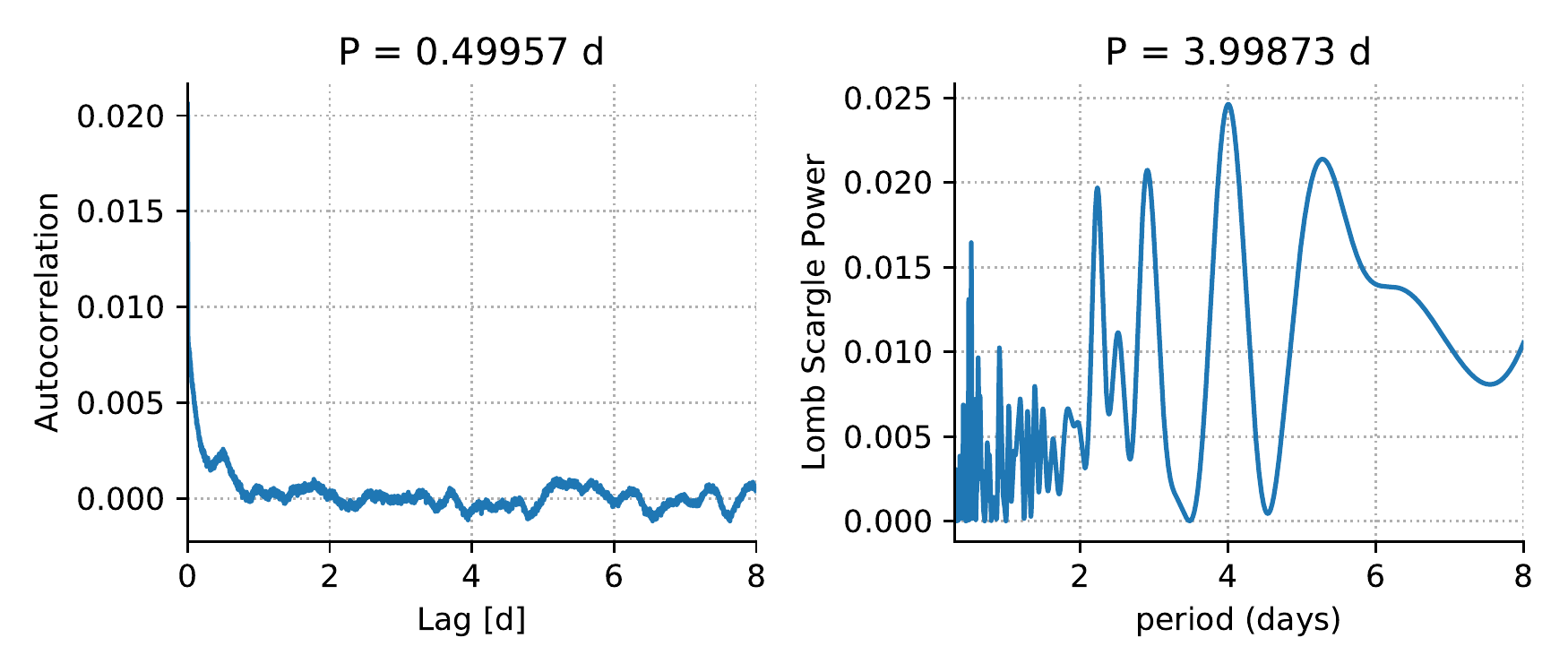}
\end{center}
\caption{\textsl{Upper:} \spitzer observations of TRAPPIST-1  after transits and flares have been removed (black circles) \citep{Delrez2018}, and a fit with a Gaussian process assuming a simple-harmonic oscillator kernel (blue curve). \textsl{Lower:} The autocorrelation function and Lomb-Scargle periodogram for the \spitzer observations show a characteristic peak at periods near 0.5 days. \label{fig:spitzer}}
\end{figure}

One might expect that the periodicity observed with \kepler at 3.3 days would be present in \spitzer observations at 4.5 $\mu$m, but after removing flares and transits from the \spitzer light curve, little signal of rotation is present \citet{Delrez2018}, see Figure~\ref{fig:spitzer}. We compute the autocorrelation function and Lomb-Scargle periodogram for the \spitzer light curve and find one periodic component near 0.5 days -- similar to the timescale of super-granulation on the Sun \citep{aigrain2003} -- in addition to periodicity near 4 days. 

To measure the significance of the peaks in the Lomb-Scargle periodogram, we generate 100 simulated light curves -- each light curve was a Gaussian process sample drawn from the maximum-likelihood kernel fit to the Spitzer observations -- and found that 33\% of the random light curves produced LS peaks with at least as much power as the 4 d peak, suggesting that all signals in the LS periodogram are insignificant.

\begin{figure*}
\begin{center}
\includegraphics[scale=0.7]{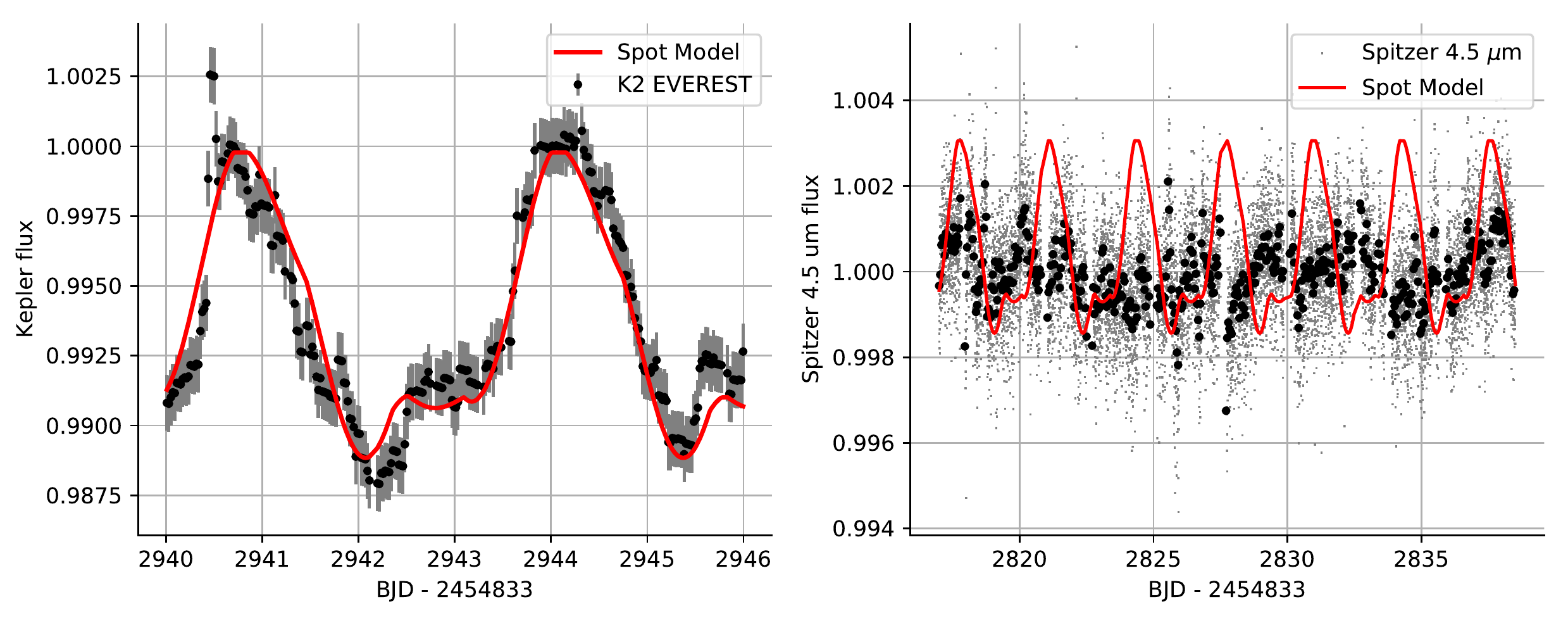}
\end{center}
\caption{\textsl{Left:} Maximum-likelihood spot model fit (red curve) to a segment of the K2 light curve (black points) assuming that the rotational modulation is driven by dark spots. \textsl{Right:} the same spot modulation extrapolated out to the Spitzer 4.5 $\mu$m band (red curve). The gray dots show the \spitzer observations, the black circles are one-hour bins. The \spitzer light curve does not show this variability, suggesting that dark spots may not be driving the photometric variability in the \kepler band. \label{fig:dark_spots}}
\end{figure*}

We choose a small section of the \kepler/K2 EVEREST light curve to fit with the spot model, see Figure~\ref{fig:dark_spots} \citep{luger2017everest}. We begin by normalizing the light curve by a quadratic, and median-filter the fluxes over a five-cadence kernel to remove most flares. We select a portion of the light curve (black points) that has a repeated pattern across more than one stellar rotation. There are very few repeated flux patterns in the \kepler light curve, indicating that the spot evolution timescale is likely shorter than the rotation period. We construct a spot model with three spots and allow their positions, radii and contrast to vary, with the contrast bounded on $0 < c < 1$. The blue curves in Figure~\ref{fig:dark_spots} are models drawn from the posterior samples, indicating that the three spots have radii near $R_\mathrm{spot}/R_\star \sim 0.12$ and contrast $c = 0.6$. 

The prior applied to ensure that starspots are dark ($c<1$) is informed by our studies of the Sun and sunlike stars, which have dark starspots \citep[see e.g.][]{Solanki2003, Morris2017a}.  However, bright spots have been observed on nearby brown dwarf the Luhman 16A via Doppler imaging \citep{Crossfield2014}, and on the \textit{T Tauri} star LkCa 4 \citep{Gully2017}.

\begin{figure}
\begin{center}
\includegraphics[scale=0.85]{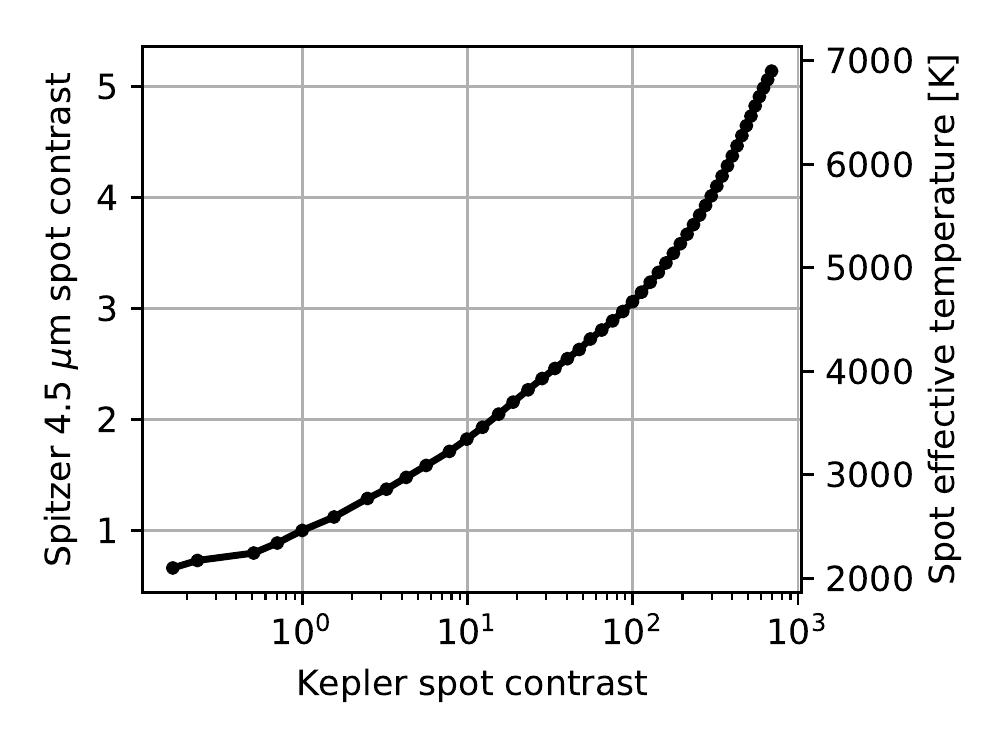}
\end{center}
\caption{Equivalent spot contrasts in the \spitzer and \kepler bandpasses, and the spot temperatures that produce those contrasts, for a star with $T_\mathrm{eff} = 2500$ K using PHOENIX+BT Settl model spectra, as described in Equation~\ref{eqn:contrast}. \label{fig:contrast}}
\end{figure}

To investigate the theoretical effects of bright spots or dark spots, we compute the contrast of a spot with spectrum $\mathcal{F_\mathrm{spot}}$ on a star with spectrum $\mathcal{F_\mathrm{photosphere}}$, using spectra drawn from the  PHOENIX+BT Settl model \citep{Husser2013}. We find that
\begin{equation}
c = \frac{{\displaystyle \int} \mathcal{F_\mathrm{\lambda, spot}} \mathcal{T}_\lambda \lambda ~  d\lambda}{{\displaystyle \int} \mathcal{F_\mathrm{\lambda, photosphere}} \mathcal{T}_\lambda \lambda ~ d\lambda} \label{eqn:contrast}
\end{equation}
where $\lambda$ is wavelength, and $\mathcal{T}$ is the filter transmission curve for \kepler or \spitzer. We compute $c$ for spots on a star with $T_\mathrm{eff} = 2500$ K in Figure~\ref{fig:contrast}. For \kepler contrasts $0 < c < 1$, the \spitzer contrast is $0.5 < c < 1$ --- in other words, the spot contrast of dark spots in the \kepler band is similar to the spot contrast of dark spots in the \spitzer band. This result is in contradiction with the observed spot modulation in the \kepler bandpass and the undetected modulation in the \spitzer light curve.

Alternatively, the contrast of \textit{bright} spots in the \spitzer 4.5 $\mu$m band increases very slowly as the spot contrast in the \kepler bandpass exceeds $c > 1$ (Figure~\ref{fig:contrast}). A spot 100x brighter than the photosphere in the \kepler band would only be 3x brighter in the \spitzer band. This slow increase in the contrast of bright spots at long wavelengths could produce large flux modulations in the \kepler bandpass and very small modulations in the \spitzer bandpass for very small spots. 

\begin{figure}
\begin{center}
\includegraphics[scale=0.6]{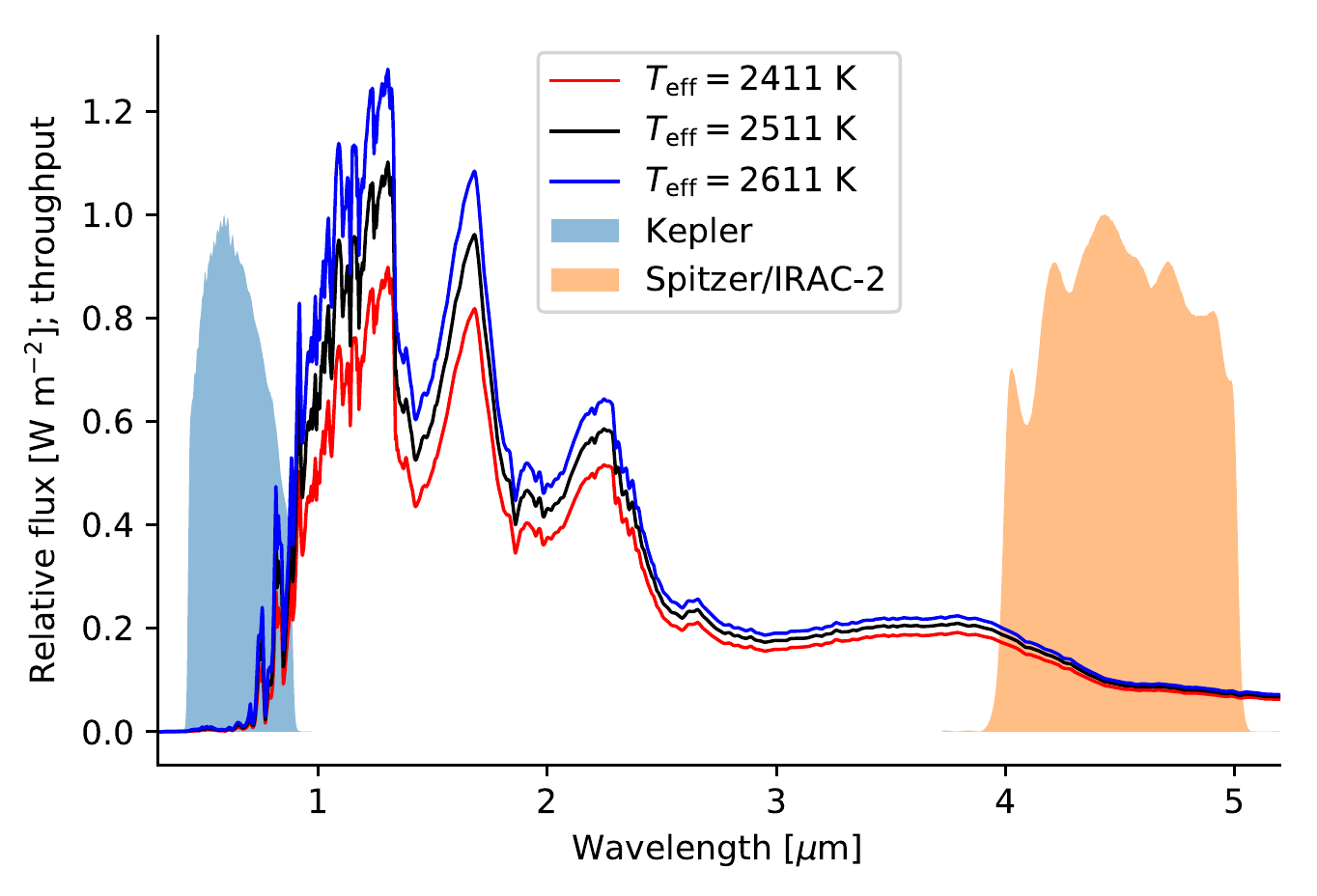}
\end{center}
\caption{Comparison of the \spitzer and \kepler bandpasses with a PHOENIX model atmosphere with $T_\mathrm{eff} = 2511$ K (TRAPPIST-1 has $T_\mathrm{eff} = 2511 \pm 37$ K, see \citealt{Delrez2018}). \label{fig:bandpasses}}
\end{figure}

\section{Simultaneous fits to the \kepler and \spitzer light curves} \label{lightcurve_fits}

We fit the spot model to the \kepler light curve and a segment of the \spitzer observations of the same duration, using the empirical relation from Figure~\ref{fig:contrast} to convert spot contrasts in the \kepler bandpass to contrasts at 4.5 $\mu$m. We fit both light curves simultaneously, even though the observations were not simultaneous, so that the \spitzer light curve will constrain the spot radii and contrasts to be consistent with the typical variability in the infrared. The \kepler observations ended 76 days before the \spitzer observations started. We assume that the variability of the star did not change significantly over those 76 days because: (1) the variability does not seem to change significantly throughout the duration of either the \kepler or \spitzer light curves, and (2) the elapsed time between observations is short compared to the length of activity cycles of M stars (on the order of years -- \citealt[see e.g.~][]{Mascareno2016}).


We must first choose the minimum number of spots required to reproduce the K2 light curve. We fit the K2 light curve for $n={1, 2, ..., 6}$ spots, allowing the spot positions, radii, and contrast to vary, and evaluate the reduced $\chi^2$ and total spotted area of each fit. We find that the reduced $\chi^2$ plateaus near a minimum after 3 or more spots have been added. If more than three spots are modelled, the radii of the spots are decreased in order to keep the total spotted area approximately constant. We therefore fix our spot number to 3 spots since that minimizes the $\chi^2$ and the number of free parameters. 

If there are many small spots distributed isotropically on the stellar surface, we would not detect them through rotational modulation. As a result, the rotational modulation fits are only sensitive to the longitudinal asymmetries in the spot distribution. Consequently, the spotted areas inferred from this model should be considered lower limits on the spotted area on the star.

We impose a prior to ensure that the addition of bright, hot spots does not change the color of TRAPPIST-1 significantly. We measure the color of the star from the optical spectrum $V-I_c = 4.7$ \citep{Burgasser2015}. At each step in our Markov chains, we add to the prior a penalty for significant color deviations from the observed color, 
\begin{equation*}
\log p \propto -\frac{1}{2}\left[ (V-I_c)_\mathrm{model} - (V-I_c)_\mathrm{observed} \right]^2.
\end{equation*}

\begin{figure*}
\begin{center}
\includegraphics[scale=0.7]{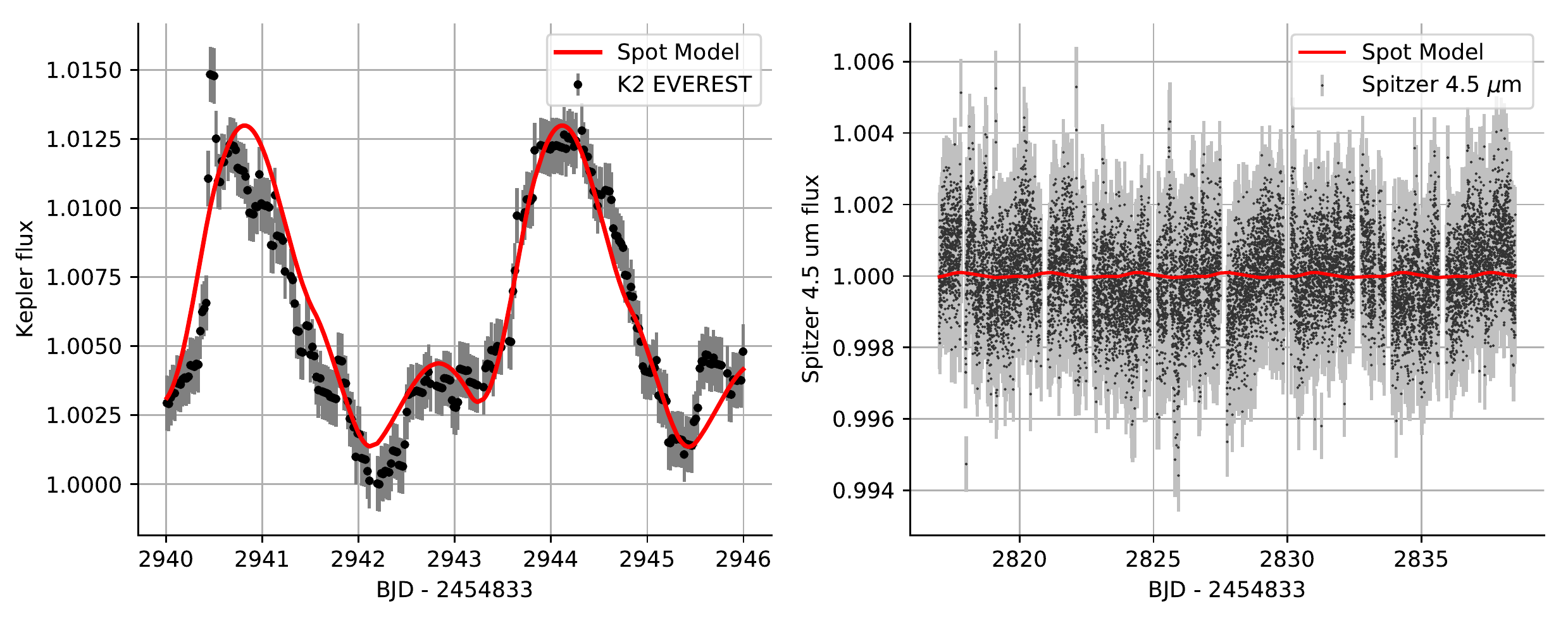}
\end{center}
\caption{Maximum-likelihood model fit for bright spots on TRAPPIST-1 (red curve), fitting the \kepler (left) and \spitzer (right) light curves (black points) simultaneously. \label{fig:fits}}
\end{figure*}

The maximum-likelihood bright spot model fit to the \kepler and \spitzer light curves\footnote{See Figure~\ref{fig:corner} in the Appendix for the complete posterior sample corner plot.} is shown in Figure~\ref{fig:fits}. The relative flux is normalized to its minimum, which assumes that the minimum flux within this segment of the light curve represents the unspotted flux. The variability in the \kepler band is reproduced by the spot model, and the corresponding variability in the \spitzer band is comparable to the observational uncertainties. 

One spot has radius $R_\mathrm{spot}/R_\star = 0.02 \pm 0.002$ and the other two have $R_\mathrm{spot}/R_\star = 0.013 \pm 0.002$. We measure the \kepler contrast $c_k = 230 \pm 40$, which corresponds to \spitzer 4.5 $\mu$m contrast of $c_s = 3.7 \pm 0.1$, and a spot temperature of $T_\mathrm{spot} \gtrsim 5300 \pm 200$. The uncertainty in the minimum spot temperature is likely underestimated, since the Markov chains prefer a narrow range of spot contrasts to fit the \kepler light curve, but the \spitzer light curve weakly constrains the lower limit on the spot temperature. The total bright spot area coverage is 16 microhemispheres (one hemisphere is half the surface area of the star) -- which is small compared to the typical dark spot area coverage on the Sun \citep{Morris2017a}. 

We repeat this analysis using a different segment of the K2 light curve to ensure that the results are reproducible at different times throughout the rotation of TRAPPIST-1. We choose the fluxes over two rotations spanning $2457762 < \mathrm{BJD} < 2457769$ and find spot sizes and contrasts consistent with the results from the other segment of the K2 light curve (both regions are labeled on Figure~\ref{fig:flare_analysis}). 

\begin{figure}
\begin{center}
\includegraphics[scale=0.8]{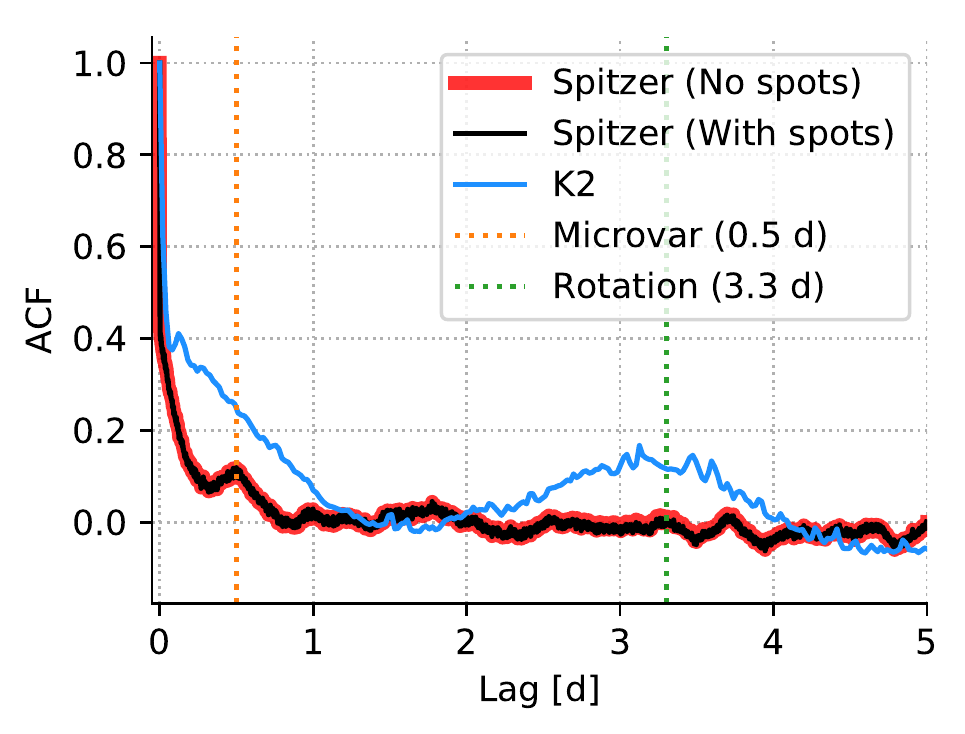}
\end{center}
\caption{The autocorrelation function of the \spitzer observations (red), the \spitzer observations plus the maximum-likelihood spot model (black), and the K2 observations (blue), with special timescales noted for microvariability and stellar rotation. \label{fig:acf}}
\end{figure}

The autocorrelation function of the \spitzer observations plus the maximum-likelihood spot model serves as a sanity-check on the amplitude of the signal introduced by the inferred bright spots. If injecting the maximum-likelihood spot model into the \spitzer light curve introduces significant periodicity, we should see an uptick in the autocorrelation function at $P_\mathrm{rot} = 3.3$ d. Shown in Figure~\ref{fig:acf}, the autocorrelation function is relatively unchanged by the injection of the spot modulation from the maximum-likelihood spot model, and which suggests that this spot model is plausibly consistent with the \spitzer observations. 

\section{Correlation between bright spots and flares} \label{correlation}

The flares of active M4 dwarf GJ 1243 have been studied in detail \citep{Hawley2014, Davenport2014}. The authors searched for a correlation between flare occurrence and starspot phase by comparing the quiescent flux of the star just before a flare to the mean flux. If flares occur near starspots, the quiescent flux of the star just before the flare should be less than the mean. No such correlation emerged, which the authors suggest indicates that the positions of flares and spots are uncorrelated. They also searched for correlation with rotational phase, which could be connected to long-lived polar spots, and found no correlation.

\begin{figure}
\begin{center}
\includegraphics[scale=0.45]{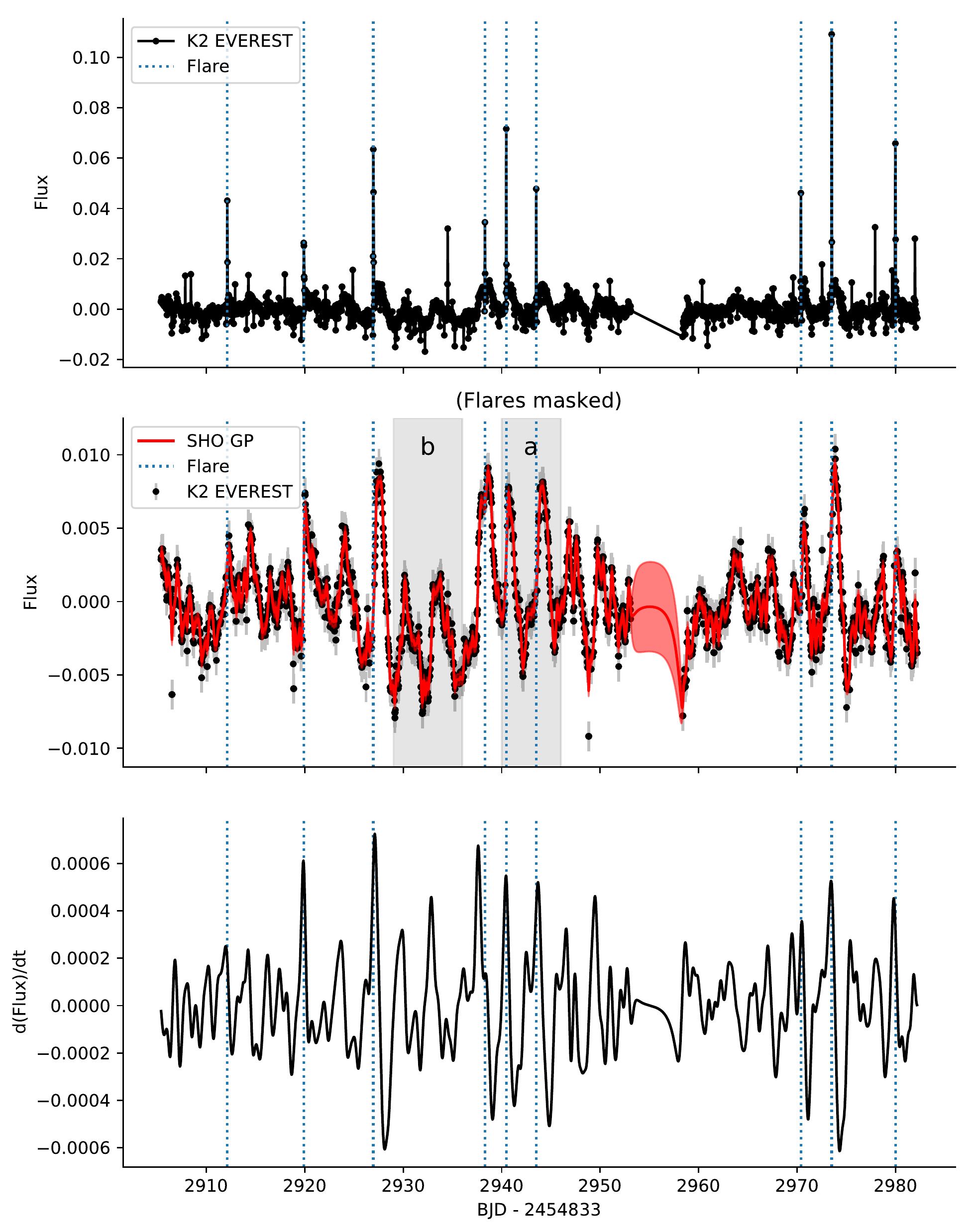}
\end{center}
\caption{\textsl{Upper}: We manually flag flares (blue dashed lines) in the K2 EVEREST light curve (black curve). \textsl{Middle:} We investigate whether or not they preferentially occur at times of high flux by masking out flares and inferring the flux at the time of the flare with a Gaussian process model using a simple harmonic oscillator kernel \citep{Foreman-Mackey2017}. We find that flares occur preferentially at times of high flux (see Figure~\ref{fig:flare_dist}). The region labeled ``a'' is the segment of the light curve in Figures~\ref{fig:dark_spots} and \ref{fig:fits}, the region labeled ``b'' is the validation segment where we repeated the bright spot analysis. \textsl{Lower}: It also appears that the flares occur on the leading edge of the brightening events. We show the numerical derivative of the Gaussian process model in the middle panel, and find that the flares tend to occur when the flux is increasing most rapidly (see also Figure~\ref{fig:folded}). \label{fig:flare_analysis}}
\end{figure}

We carried out a similar analysis to investigate if the flares are correlated with starspot phase for TRAPPIST-1, as was briefly noted in \citet{Vida2017}. We manually identified the nine largest flares in the K2 EVEREST light curve, and masked them out -- see Figure~\ref{fig:flare_analysis}. We removed a fifth order polynomial to remove systematic trends without removing stellar variability. To measure the flux of the star at the time of the flares, we fit the rotational modulation with a Gaussian process using a simple harmonic oscillator kernel, and extrapolate the model to the times of flares. 

\begin{figure}
\begin{center}
\includegraphics[scale=0.6]{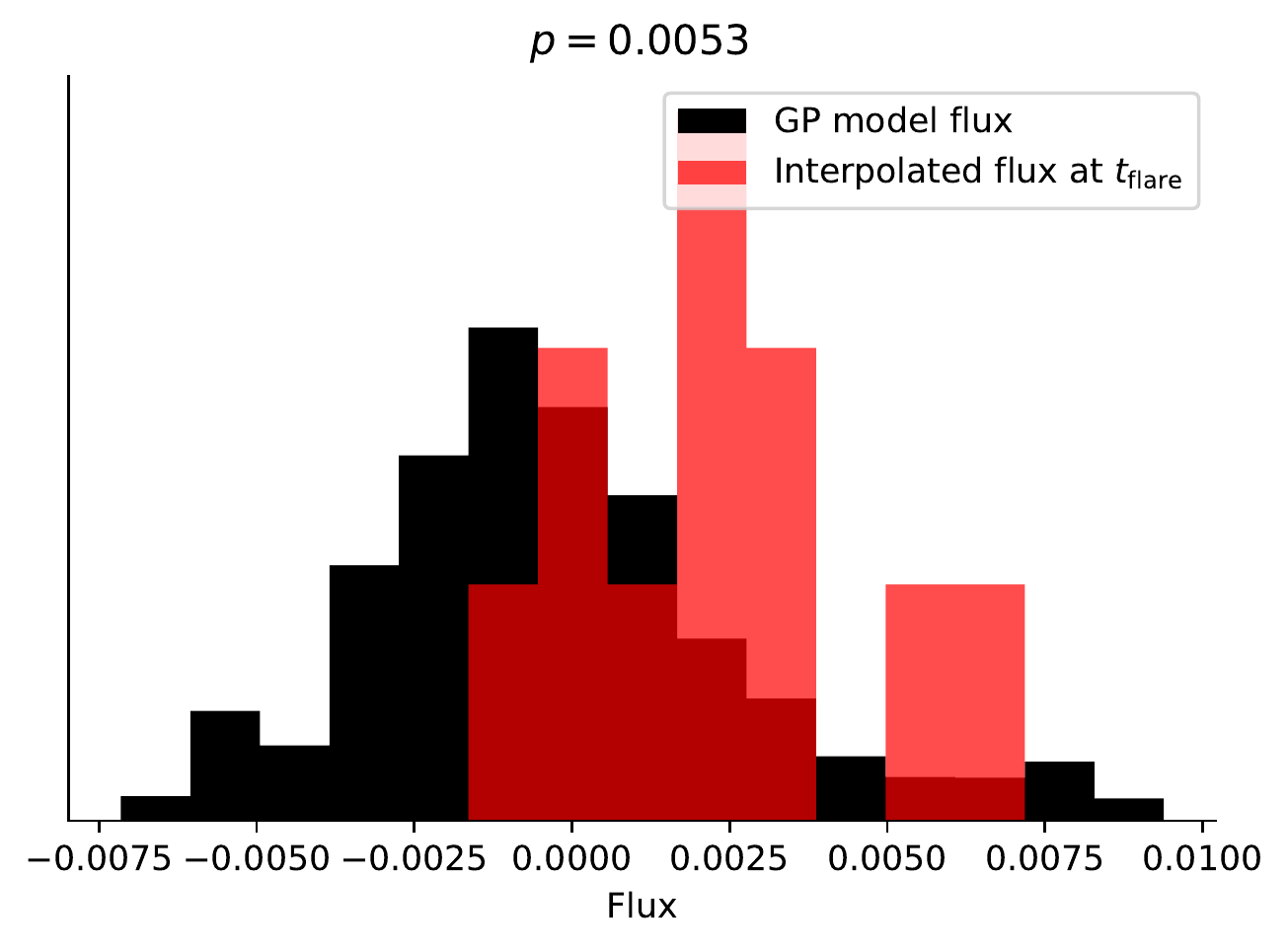}
\end{center}
\caption{Distribution of fluxes with flares masked (black), compared with the distribution of fluxes at the times of flares after flares have been masked out and interpolated with a Gaussian process model (red). The flares typically occur when the star is bright. \label{fig:flare_dist}}
\end{figure}

We find that the stellar flux at the time of flares is greater than the typical flux, see Figure~\ref{fig:flare_dist}. The two sample T-test yields $p = 0.005$ for the two flux distributions, indicating some significance to the difference in mean fluxes. In other words, the star is typically brighter just before a flare event than the mean flux. This might suggest that the bright active regions are spatially correlated with the flares on TRAPPIST-1. 

The flares also tend to occur when the change in brightness is most positive (see Figure~\ref{fig:flare_analysis}). If we interpret the 3.3 day periodicity as rotational modulation, then the preference for flares to occur on the leading edge of the brightening events would indicate that flares are most likely to occur at a particular stellar longitude, in contradiction with the flare analysis of \citet{Vida2017}. If we instead interpret the 3.3 day periodicity as the characteristic lifetimes of bright active regions on the star, then the 1\% brightening events may be localized brightening associated in time and space with the flaring activity.  Until a robust measurement of the stellar rotation period has been made, we cannot rule out the possibility that the 3.3 d period is not the stellar rotation period.

\begin{figure}
\begin{center}
\includegraphics[scale=0.9]{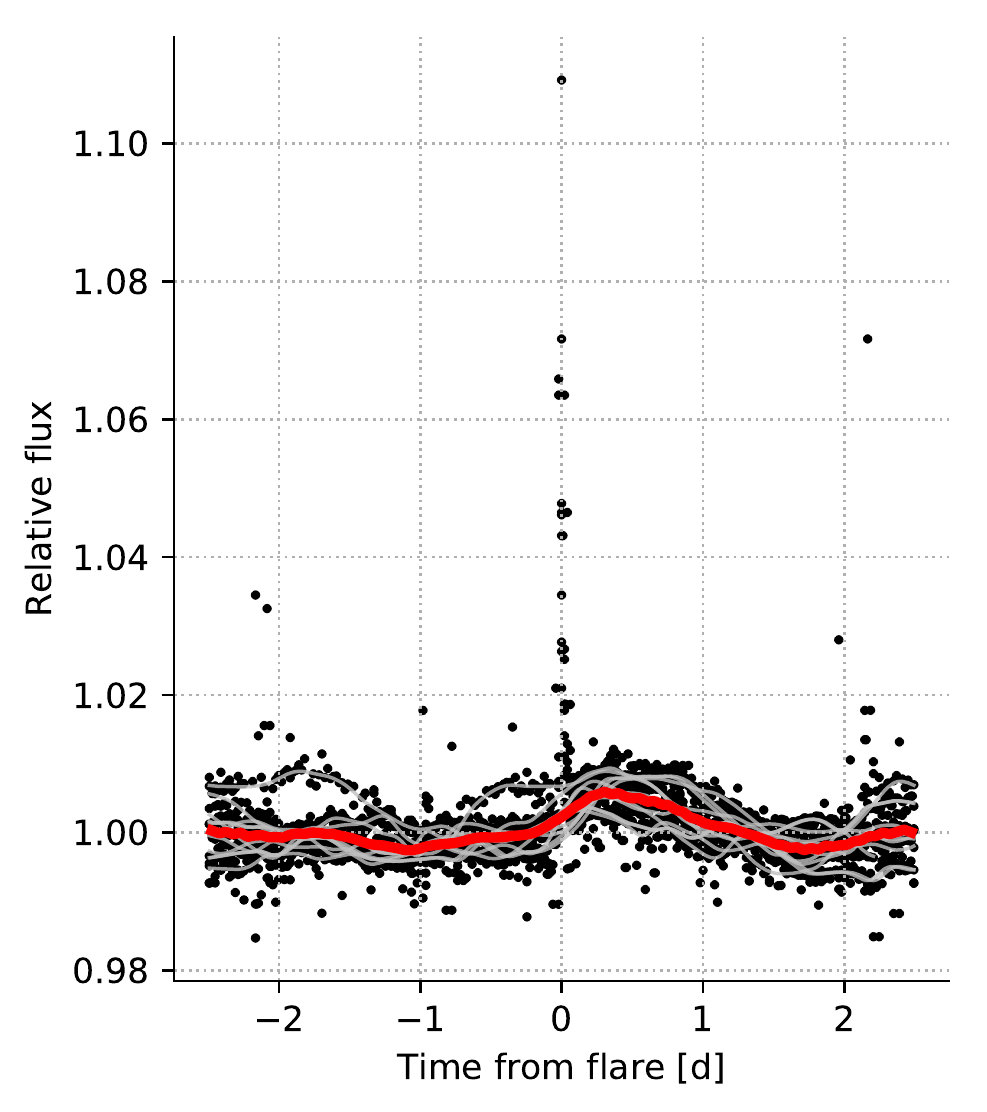}
\end{center}
\caption{K2 light curve (black circles) and the maximum-likelihood Gaussian process models (gray curves), folded at the times of flares. The red curve is the mean of the Gaussian process models. It appears that the flares occur preferentially during the rise in brightness (see also Figure~\ref{fig:flare_analysis}). \label{fig:folded}}
\end{figure}

\section{Discussion}

Bright spots on the Sun referred to as faculae arise due to the magnetic fields and viewing geometry of convective granules \citep{Spruit1976, Spruit1977}. Solar faculae have sizes comparable to convective cells \citep{Keller2004}, and intensity contrasts relative to the mean photosphere that are small ($c \sim1.5$) compared to the spot contrast we observe on TRAPPIST-1 in the \kepler band. The characteristic granule size on late-M dwarfs is likely to be $\sim80$ km across \citep{ludwig2002}, while the spots on TRAPPIST-1 are $\sim 600$ km in diameter. We suggest that the spots on TRAPPIST-1 should not be considered faculae.

Few observations have revealed small scale active regions on fully-convective stars. The global-scale magnetic fields of late-M dwarfs have been studied in detail using Zeeman-Doppler Imaging \citep{Donati2003, Morin2010, Morin2011, Morin2013}. One exception is the fully convective star V374 Peg which shows 2\% dark spot coverage \citep{Morin2008}. 

We can compare our results with \citet{Rackham2017b} and \citet{Zhang2018}, who computed the spot and faculae covering fractions for TRAPPIST-1. Both studies find that the stellar spectrum is best described by a heterogeneous surface, with persistent spectral components at different temperatures. Our constraints, based upon two-band photometry, also imply that the star is best described as a heterogeneous mixture, described by bright spots and a dimmer photospheric component. We are proposing bright spots with higher temperatures, smaller covering factor, and shorter lifetimes than the steady spotted and facular components of \citet{Rackham2017b} or \citet{Zhang2018}.  It is possible that the star has some combination of all of these features (bright spots, dark spots, and faculae).

In the previous section, we proposed that the apparent time-correlation between the occurrence of luminous flares and the brightness of the star is due to a physical association between the positions of bright spots which are associated with flares. An alternative hypothesis is that the $\sim 1$\% flux variations are not rotational modulation, but rather transient, bright active regions which accompany flares. If that is the case, then the 3.3 d periodicity in the K2 light curve should be interpreted as a characteristic active region timescale rather than a rotation period.  In fact, if the bright spots were due to rotational variability, this might imply that the luminous flares preferentially occur at the same stellar longitude.  This seems implausible given that magnetic activity on the surface of the star shouldn't be connected to the inertial frame.  Thus, we feel that the possible correlation between the bright flares and spots may argue against the spot variation being due to rotation.

Hydrogen Balmer line emission (H$\alpha$) is common in late M dwarfs and variable H$\alpha$ emission could potentially explain the 3.3-day modulations observed by \kepler; as H$\alpha$ is not within the \spitzer bandpass, this would naturally explain the lack of a 3.3-day feature in the \spitzer lightcurve.  The optical spectrum of TRAPPIST-1 of \citet{Burgasser2015} shows that the flux in H$\alpha$ is only 0.3\% of the flux integrated over the Kepler bandpass. However, \citet{Kruse2010} found that H$\alpha$ equivalent widths vary by up to a factor of five for M8V stars, indicating that it is plausible that extreme H$\alpha$ variability (by a factor of $\sim 5$) could explain the \kepler variability. Further observations of H$\alpha$ emission from TRAPPIST-1 are required to investigate this possibility.

\section{Conclusions}

We simultaneously model photometry of TRAPPIST-1 from \kepler and \spitzer to measure the properties of its putative starspots. We find that if the 3.3 day periodicity is due to stellar rotation, TRAPPIST-1 likely has a few bright spots rather than dark spots, and the spots have characteristic temperatures $T_{eff} \gtrsim 5300$ K and radii $R_\mathrm{spot}/R_\star \sim 0.004$. 

The bright spots add a source of flux dilution to the transit light curves of each planet. We provide a correction factor for the transit depths of each planet, and propagate those depths to revise planet radii and densities; however, we note that other sources of variability and/or stellar inhomogeneity likely dominate this correction in the infrared \citep{Delrez2018,Zhang2018}.

We note that flares occur preferentially when the star is bright, and when the brightness is increasing most rapidly. This may suggest that that the flares are associated with the hot spots. Alternatively, the brightness variations could be the growth and decay of bright active regions on the stellar surface with a characteristic timescale of 3.3 days.

Though the nature of the proposed stellar activity is still unclear, the observations suggest that TRAPPIST-1 has bright spots rather than dark ones. Even if the continual spot variability observed by K2 were due to transient photospheric spots, rather than stellar rotation, the rapid appearance and disappearance of \textit{dark} photospheric spots would produce a signal which was not observed by Spitzer. 

\software{\texttt{ipython} \citep{ipython}, \texttt{jupyter} \citep{Kluyver2016}, \texttt{numpy} \citep{VanDerWalt2011}, \texttt{scipy} \citep{scipy},  \texttt{matplotlib} \citep{matplotlib}, \texttt{astropy} \citep{Astropy2013}, \texttt{gatspy} \citep{gatspy}, \texttt{emcee} \citep{Foreman-Mackey2013}, \texttt{celerite} \citep{Foreman-Mackey2017}}

\facility{Spitzer, \kepler/K2}

\acknowledgements

We gratefully acknowledge Laetitia Delrez, Brice Demory and Michael Gillon for sharing their detrended \spitzer observations, and we thank Ethan Kruse and Pey Lian Lim for productive conversations.  This work is based in part on observations made with the Spitzer Space Telescope, which is operated by the Jet Propulsion Laboratory, California Institute of Technology under a contract with NASA. Support for this work was provided by NASA through an award issued by JPL/Caltech. 

We acknowledge support from NSF grant AST-1615315, NASA grant NNX14AK26G, and from the NASA Astrobiology Institute's Virtual Planetary Laboratory Lead Team, funded through the NASA Astrobiology Institute under solicitation NNH12ZDA002C and Cooperative Agreement Number NNA13AA93A

This work made use of \texttt{synphot} (Lim, P. L., et al. 2016, synphot User's Guide, STScI, \url{http://synphot.readthedocs.io/en/latest/}). This research has made use of NASA's Astrophysics Data System. This research has made use of the SVO Filter Profile Service (\url{http://svo2.cab.inta-csic.es/theory/fps/}) supported from the Spanish MINECO through grant AyA2014-55216 \citep{rodrigo2012}.

\appendix

\section{The effect of bright spots on transit light curves} \label{sec:transits}

Typically the transit depth is assumed to be the ratio of the cross-sectional areas of the projected planet on the star, 
\begin{equation}
\delta_\mathrm{unspotted} = \frac{\pi R_p^2}{\pi R_\star^2},
\end{equation}
where we use the equality sign because we are ignoring limb-darkening in this example. If there are bright unocculted spots on the star, the measured depth will be 
\begin{equation}
\delta_\mathrm{spotted} = \frac{\pi R_p^2}{\pi R_\star^2 + (c-1) \pi R_\mathrm{spot}^2}
\end{equation}
where $c$ is the spot contrast relative to the photosphere. Rearranging, we find
\begin{equation}
\frac{\delta_\mathrm{unspotted}}{\delta_\mathrm{spotted}} = 1 + (c-1) (R_\mathrm{spot}/R_\star)^2.
\end{equation}
This correction for the unocculted bright spots in the \kepler and \spitzer bands are $\delta_\mathrm{unspotted}/\delta_\mathrm{spotted} = 1.004 \pm 0.001$ and $1.00006 \pm 0.00001$, respectively. These transit depth dilution corrections allow us to update the observed planet radii reported by \citet{Gillon2017} with K2, which we list in Table~\ref{tab:planets}.  If the are of bright spots is confined to the three spots we modeled, then the dominant systematic affecting planet radii from the \spitzer observations is the microvariability observed in Figure~\ref{fig:spitzer}, rather than the bright spot variability.   However, we note that this may be a lower limit on the effect of bright spots as our three-spot model only measures the {\it variable} component of the bright spots.  We have carried out a test in which we added numerous small spots distributed in longitude, which produces an equally good fit to the variable light curve, but with much larger areal coverage of bright spots.

\begin{figure}
\begin{center}
\includegraphics[scale=0.7]{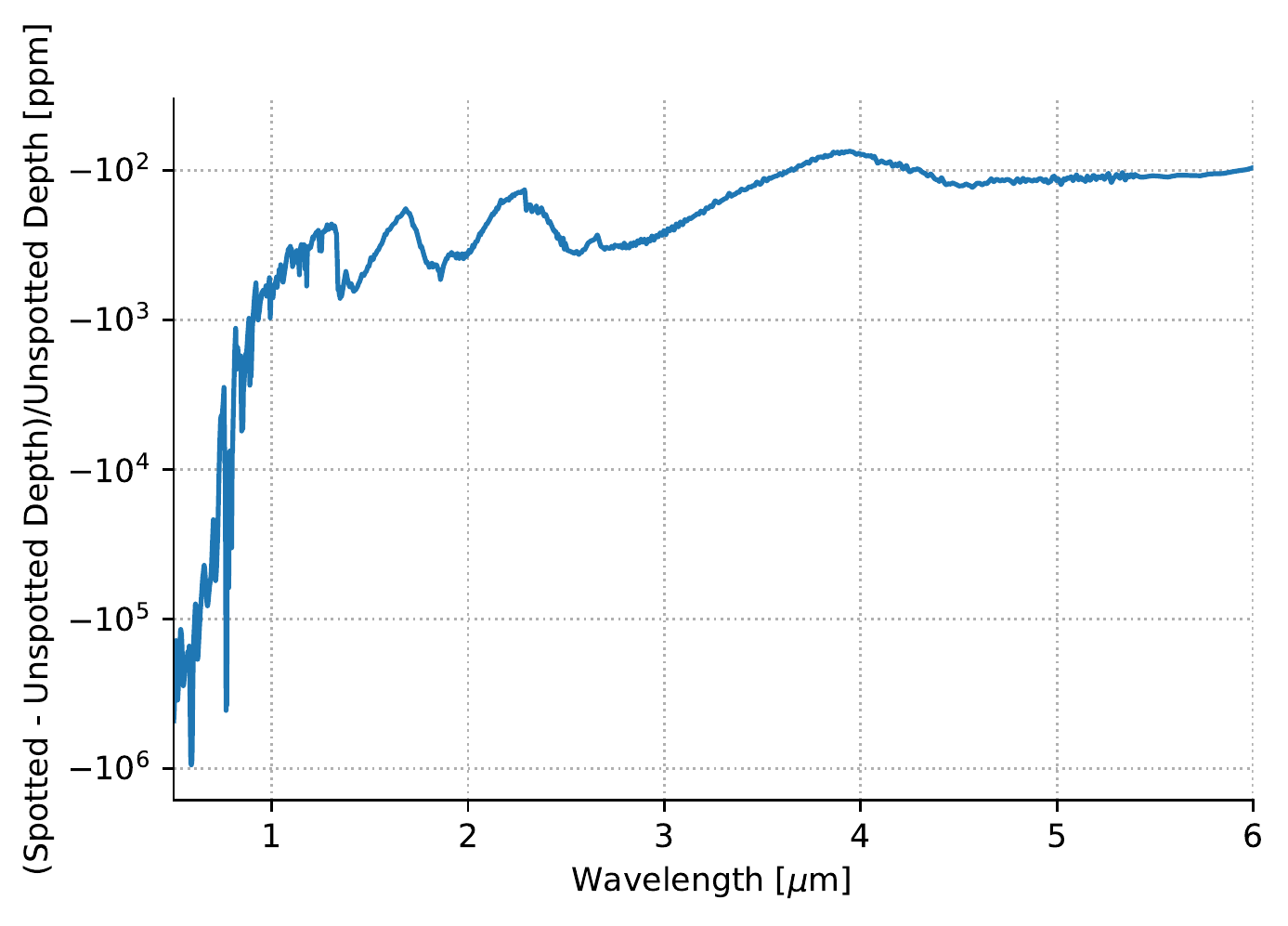}
\end{center}
\caption{Fractional change in transit depth in the spotted vs.\ unspotted case.  \label{fig:transmission}}
\end{figure}

The flux dilution due to the bright spots on TRAPPIST-1 will cause the transit depths to appear a shallower than they truly are. In Table~\ref{tab:planets} we list the revised planet properties -- all revisions are within the uncertainties of the measurements. The effect as a function of wavelength is plotted in Figure~\ref{fig:transmission}.

\begin{table*}
\centering
\begin{tabular}{l|cccccc}
 & Depth (\%) & Depth (\%) & Radius [$R_\oplus$] & Radius [$R_\oplus$] & Density [$\rho_\oplus$]& Density [$\rho_\oplus$] \\
& (G17) & (this work) & (G17) & (this work)  & (G17) & (this work) \\ \hline
b & $0.727 \pm 0.009$ & $0.729 \pm 0.009$ & $1.086 \pm 0.035$ & $1.090 \pm 0.034$ & $0.660 \pm 0.560$ & $0.656 \pm 0.556$ \\
c & $0.687 \pm 0.010$ & $0.690 \pm 0.010$ & $1.056 \pm 0.035$ & $1.060 \pm 0.034$ & $1.170 \pm 0.530$ & $1.159 \pm 0.512$ \\
d & $0.367 \pm 0.017$ & $0.368 \pm 0.017$ & $0.772 \pm 0.030$ & $0.775 \pm 0.030$ & $0.890 \pm 0.600$ & $0.882 \pm 0.581$ \\
e & $0.519 \pm 0.026$ & $0.521 \pm 0.026$ & $0.918 \pm 0.039$ & $0.921 \pm 0.037$ & $0.800 \pm 0.760$ & $0.793 \pm 0.742$ \\
f & $0.673 \pm 0.023$ & $0.676 \pm 0.023$ & $1.045 \pm 0.038$ & $1.049 \pm 0.037$ & $0.600 \pm 0.170$ & $0.589 \pm 0.156$ \\
g & $0.782 \pm 0.027$ & $0.785 \pm 0.027$ & $1.127 \pm 0.041$ & $1.131 \pm 0.040$ & $0.940 \pm 0.630$ & $0.927 \pm 0.609$ \\
h & $0.352 \pm 0.033$ & $0.353 \pm 0.033$ & $0.755 \pm 0.034$ & $0.759 \pm 0.042$ & --- & --- \\
\end{tabular}
\caption{Revised planet properties in the \kepler bandpass accounting for the flux dilution due to bright spots on TRAPPIST-1. Here we take planet masses and the definition of Depth = $(R_p/R_\star)^2$ as in \citet{Gillon2017}. \label{tab:planets}}
\end{table*}

\section{Posterior distributions}

\begin{figure*}
\begin{center}
\includegraphics[scale=0.3]{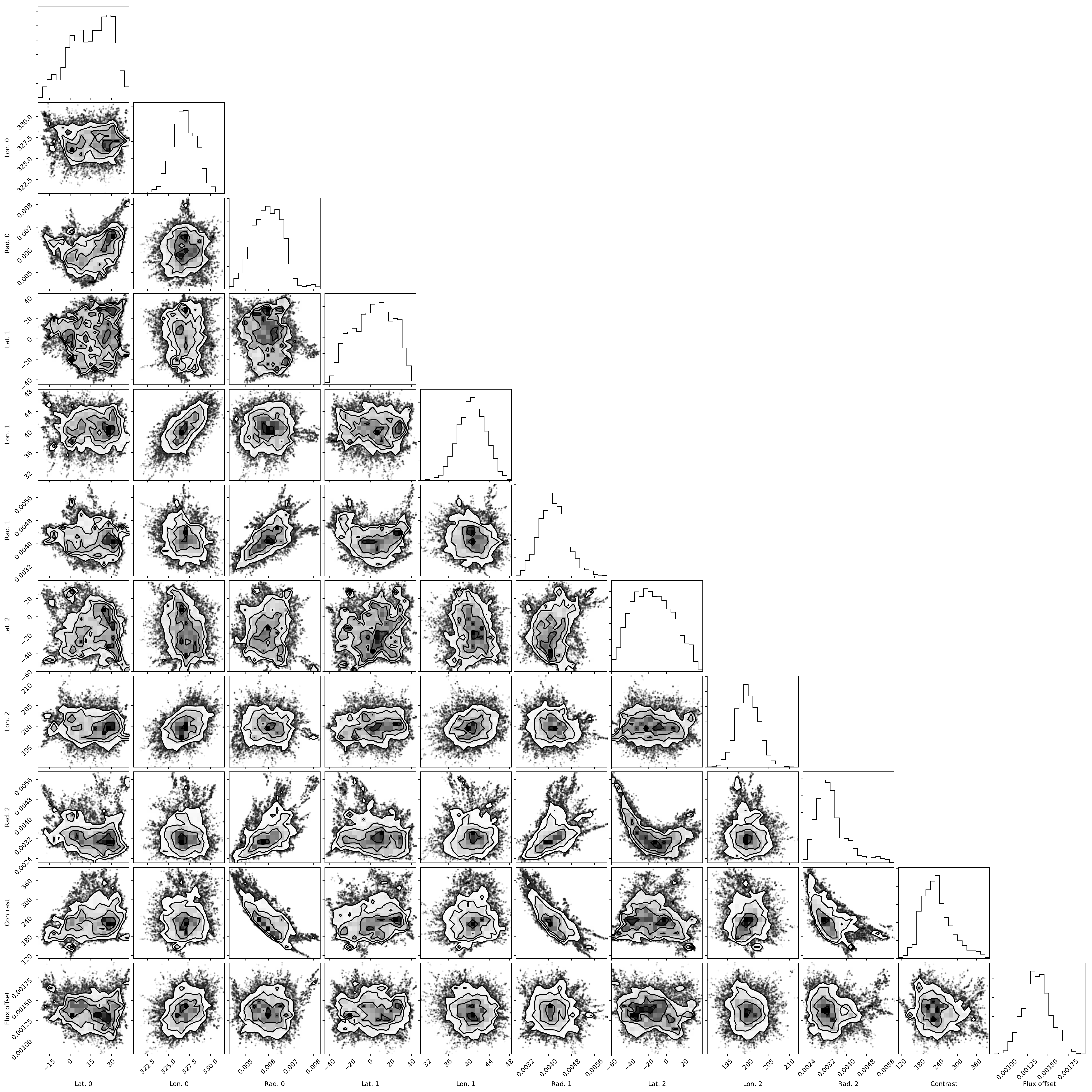}
\end{center}
\caption{Posterior samples for the three, bright spot model. \label{fig:corner}}
\end{figure*}

The full posterior distributions on all fit parameters in the bright spot model are shown in Figure~\ref{fig:corner}. 

\section{Probability of spot occultations}

Occultations of the bright spots by the planets would allow us to infer the spot properties independently, as we showed for HAT-P-11 in \citet{Morris2017a}. We calculate an upper-limit on the probability that the proposed bright spots will be occulted by planets in the TRAPPIST-1 system. If we assume that the spots are infinitely long-lived, each spot is visible for half of the stellar rotation. We convert the posterior samples for spot latitudes to posterior samples of impact parameters $p_b$, and integrate the posterior samples for $b$ over the range of impact parameters occulted by each planet ($b_0$, $b_1$), normalized by the integral of the posterior samples over all impact parameters, 
\begin{equation}
P_\mathrm{occult} \approx \frac{\int_{b_0}^{b_1} p_b \; dp}{2 \int p_b \; dp} 
\end{equation}
The spot latitude solutions are degenerate: the spots could be on either side of the stellar equator and produce the same rotational modulation, and the planet impact parameter is similarly degenerate in that the planets could be occulting the northern or southern stellar hemisphere. As a result, we compute each probability of occultation twice: once assuming the planets occult the northern stellar hemisphere and once assuming they occult the southern hemisphere, and report the maximum spot occultation probability in Table~\ref{tab:prob_occ}. 

\begin{table}
\centering
\begin{tabular}{cc}
Planet & $P_\mathrm{occult, max}$ \\ \hline
b & 0.08 \\
c & 0.08 \\
d & 0.05 \\
e & 0.06 \\
f & 0.06\\
g & 0.08 \\
h & 0.05 \\
\end{tabular}
\caption{Maximum probability of a spot occultation for each planet in the TRAPPIST-1 system, assuming infinitely long-lived spots.\label{tab:prob_occ}}
\end{table}

The K2 light curve shows evolution of the flux modulation on timescales of days, indicating that spots are not infinitely long-lived. Thus these conservative upper limits are likely much greater than the probability of a spot occultation for transient spots on TRAPPIST-1.

\end{document}